\documentclass[preprint,10pt, review, authoryear]{elsarticle}

\usepackage{amssymb}
 \usepackage{amsthm}

\usepackage{lineno,hyperref}
\modulolinenumbers[5]

\usepackage{bm}
\usepackage{subfig}
\usepackage{amsmath}
\usepackage[titletoc,title]{appendix}

\newtheorem{theorem}{Theorem}[section]
\newtheorem{corollary}[theorem]{Corollary}
\newtheorem{lemma}[theorem]{Lemma}

\newcommand{\bX}{\bm{X}}

\newcommand{\bZ}{\bm{Z}}
\newcommand{\bV}{\bm{V}}
\newcommand{\bU}{\bm{U}}
\newcommand{\bW}{\bm{W}}
\newcommand{\balpha}{\bm{\alpha}}
\newcommand{\bgamma}{\bm{\gamma}}
\newcommand{\btheta}{\bm{\theta}}
\newcommand{\bxi}{\bm{\xi}}

\begin{document}

\begin{frontmatter}



\title{A Continuous Threshold Expectile Model 
}

 \author[label1,label2]{Feipeng Zhang}
 \address[label1]{Department of Statistics, Hunan University, Changsha, 410082, China}

\author[label2]{Qunhua Li\corref{cor1}}
\address[label2]{Department of Statistics, Pennsylvania State University,  PA, 16802, USA}
\cortext[cor1]{Department of Statistics, Pennsylvania State University,  PA, 16802, USA}
\ead{qunhua.li@psu.edu}

\begin{abstract}

Expectile regression is a useful tool for exploring the relation between the response and the explanatory variables beyond the conditional mean. 
This article develops a continuous threshold expectile regression  for modeling data in which 
the effect of a covariate on the response variable is linear but varies below and above an  unknown threshold in a continuous way. 
Based on a  grid search approach, we obtain estimators for the threshold and the regression coefficients via an asymmetric least squares regression method. 
We derive the asymptotic properties for all the estimators and show that the estimator for the threshold achieves root-n consistency.  
We also develop a  weighted CUSUM type test statistic for the existence of a threshold in a given expectile, and derive its asymptotic properties under both the null and the local alternative models.  
This test only requires fitting the model under the null hypothesis in the absence of a threshold, 
thus it is computationally more efficient than the likelihood-ratio type tests. 
Simulation studies show desirable finite sample performance in both homoscedastic and heteroscedastic cases.  
The application of our methods on a Dutch growth data and a baseball pitcher salary data reveals interesting insights. 

\end{abstract}

\begin{keyword}
Expectile regression \sep Threshold \sep Weighted CUSUM test 
\sep Grid search method



\end{keyword}

\end{frontmatter}

\linenumbers


\section{Introduction}
\label{s:intro}

Expectile  regression,  
first  introduced by \citet{aigner1976estimation}  and \citet{newey1987asymmetric}, 
has become  popular in the last decades.  
Analogous to quantile regression \citep{koenker1978regression}, 
expectile regression  draws a complete picture of the conditional distribution of the response variable given the covariates,  
making it a useful tool for modeling data with heterogeneous conditional distributions.  
As modeling tools,  quantile regression and expectile regression both have advantages over the other in certain aspects: 
quantile regression is more robust to outliers than expectile regression, 
whereas expectile regression is more sensitive to the extreme values in the response variable  than quantile regression.  
However, 
expectile regression has certain computational advantages over quantile regression \citep{newey1987asymmetric}.
First,  unlike quantile regression,  the loss function of expectile regression  is everywhere differentiable, thus its estimation is more straightforward and much quicker. 
Second,  the computation of  the asymptotic covariance matrix of the expectile regression estimator does not involve estimating the density  function of the errors.  
Besides the early development on linear expectile regression 
\citep{newey1987asymmetric, efron1991regression}, 
 many  nonparametric or semiparametric expectile regression have been developed in recent years, for example,  
 \citet{yao1996asymmetric}, \citet{de2009quantiles}, \citet{kuan2009assessing}, 
 \citet{schnabel2009optimal}, \citet{kneib2013beyond}, \citet{sobotka2013confidence}, 
 \citet{xie2014varying},  \citet{waltrup2015expectile},  \citet{kim2016nonlinear},  
 and among others. 
These models greatly improve  the flexibility of expectile regression for modeling nonlinear relationships.
 
However,  some natural phenomena call for nonlinear regression forms that exhibit structure changes,  sometimes in the form of two line segments with different slopes. 
For example, a child's height increases rapidly with age before and during puberty  
and then stops increasing in late teens. 
This implies that the growth curve of height may be described as two line segments with different slopes intersecting at a threshold. 
Another example arises from a study of the salaries of major league baseball players in 1987 \citep{hoaglin1995critical}. 
The data shows  a positive correlation between salaries and years of experience for less experienced pitchers but a negative correlation for more experienced pitchers. 
In these instances, besides the regression coefficients,  
the onset of the transition point is often of great research interest,  
for example, when a child reaches his/her full adult height or 
whether there is a prime time for pitchers' salaries. 
Although the existing spline-based \citep[e.g.,][]{schnabel2009optimal, kim2016nonlinear} or varying-coefficient expectile models \citep[e.g.,][]{xie2014varying} can capture the nonlinear relationship between the response variable and the predictors,  
they cannot provide information on the location of the threshold. 
This issue motivates us to consider a continuous threshold  model for expectile regression. 
Continuous threshold regression,  also called segmented regression or bent line regression,  
has been studied in the context of  least squares regression 
\citep{quandt1958estimation, 
quandt1960tests, 
hinkley1969inference,
feder1975asymptotic, 
chappell1989fitting, 
chan1998limiting, 
chiu2006bent,
hansen2015regression},  
quantile regression \citep{li2011bent}, 
and rank-based regression \citep{zhang2016robust}. 
However, no literature has investigated the continuous threshold expectile regression. 

In this article, we develop a continuous threshold expectile regression model.  
The contribution of this article is twofold. 
First, we propose a grid search method  to  estimate the unknown threshold and other regression coefficients. 
We derive the asymptotic properties for all the parameters including the threshold,  
and show that the estimator for the threshold achieves $\sqrt{n}$-consistency. 
Second, we develop a testing procedure for the existence of structural change at a given expectile, based on a weighted CUSUM type statistic. 
This test only requires fitting the model under the null hypothesis in the absence of a threshold, 
thus it is computationally efficient.   
The limiting distribution of the test statistic is also established.  
The estimation and testing procedures are implemented in R code,  
which is available from the first author by request.

The remainder of the article is organized as follows.  
In Section~\ref{s:method},  
we describe the continuous threshold expectile regression model,  
and develop a grid search method for estimating the unknown threshold and regression coefficients. A testing procedure for the structural change in a given expectile level is also proposed. 
In Section~\ref{s:simulation}, 
we conduct  simulation studies and two real data analyses. 
Section~\ref{s:conclude} provides the conclusion with possible future extensions.  
Technical proofs are presented in the Appendix.

\section{Methodology} 
\label{s:method}

\subsection{Model}
\label{ss:model}

Let $(Y_i, X_i, \bZ_i)$, $i=1,\cdots, n$, 
be a sequence of independent and identically distributed sample from the population 
$(Y, X, \bZ)$.  
We assume that $Y$ is the response variable,  $\bZ$ is a vector of covariates, 
and $X$ is a scalar variable, whose relationship with $Y$ changes at an unknown location.  
The population $\tau$-expectile of $Y$, $\nu_\tau(Y)$,  
minimizes the loss function 
$\hbox{E}\left[\rho_\tau(Y-\nu)\right]$, where 
\[
\rho_\tau (u) =\omega_\tau (u) u^2 =
 \begin{cases} 
      (1-\tau)u^2, & u\leq 0, \\
      \tau u^2, &  u>0, 
   \end{cases}
\]
is  the asymmetric squared error loss function, 
and $0<\tau<1$ is the parameter that controls the degree of loss asymmetry.   
Clearly, when $\tau=0.5$, the $\tau$-expectile   corresponds to the mean  of $Y$.   

In this paper, 
we model the conditional $\tau$-th expectile of $Y$ using  the continuous threshold model
\begin{align}\label{mod1}
	\nu_\tau(Y|X,\bZ) = \beta_0 + \beta_1 X + \beta_2 (X-t)_+ + \bgamma^\top \bZ, 
\end{align}
where $\theta_\tau =(\bxi^\top, t) ^\top$ are the unknown parameters of interest, 
$\bxi=(\beta_0, \beta_1, \beta_2, \bgamma^\top)^\top$ is the vector of parameters excluding the unknown location of the threshold or change point $t$, 
$\bgamma$ is a $p\times 1$ vector of parameters. 
Here, $a_+ = a I(a>0)$,  where $I(\cdot)$ is the indicator function. 
Clearly, the linear expectile regression is continuous on $X$ at $t$, but has different slopes on  either side of the threshold $t$. 
In other words, $\beta_1$ is the slope of the left line segment for $X\leq t$ and 
$\beta_1+\beta_2$ is the slope of the right line segment for $X > t$.

\subsection{Estimation procedure}
\label{ss:estimate}

To estimate   $\btheta_\tau =(\bxi^\top, t) ^\top$ at a given expectile $\tau$, 
we minimize  the objective function 
\begin{align}\label{loss}
	M_{n,\tau}(\btheta)=n^{-1}\sum_{i=1}^n  \rho_\tau\left(Y_i- \beta_0 - \beta_1 X_i - 
	\beta_2 (X_i-t)_+ - \bgamma^\top \bZ_i\right). 
\end{align}
However, due to the existence of the threshold $t$,  
the objective function \eqref{loss} is convex in $\bxi$ but non-convex in $t$, 
making it difficult to obtain its minimizer.  
One estimation approach is to use the grid search strategy, which is commonly used for bent line mean regression  \citep{quandt1958estimation, chappell1989fitting}.    
To proceed,  we  re-write the objective function \eqref{loss} with respect to 
$\bxi$ and $t$ as 
\begin{align}
	M_{n,\tau}(\btheta) \equiv M_{n,\tau}(\bxi, t)=
	n^{-1}\sum_{i=1}^n \rho_\tau\left(Y_i-\bxi^\top\bV_i(t)\right),
\end{align}
where $\bV_i(t)=\left(1, X_i, (X_i-t)_+, \bZ_i^\top\right)^\top$.  
The minimization can be carried out in two steps: 

(1) for each  $t \in \mathcal{T}$, where $\mathcal{T}$ is the range set of all $t$'s,  
obtain a profile estimate of $\bxi$ by  
\[
	\widehat{\bxi}(t) = \arg\min_{\bxi} M_{n,\tau}(\bxi, t). 
\]

(2) obtain the threshold $t$ as 
\[
	\widehat{t} =\arg\min_{t \in \mathcal{T}} M_{n,\tau}\left(\widehat{\bxi}(t), t\right). 
\] 
The estimate for $\btheta$ then is 
$\widehat{\btheta} = \left(\widehat{\bxi}(\widehat{t}), \widehat{t}\right)$.

\subsection{Asymptotic properties} 
Because the objective function is not  differentiable with respect to $\btheta$, 
it is impossible  to obtain the asymptotic properties of $\widehat{\btheta}$ using the standard theory.   
Here, we derive the asymptotic properties using the modern empirical processes theory. 
We first introduce some notations.  
Denote the true parameters as $\btheta_0$. 
Let $M_\tau(\btheta) = \mbox{E} \rho_\tau\left(Y-\bxi^\top \bV(t)\right)$, 
where $\bV_i(t)=\left(1, X, (X-t)_+, \bZ^\top\right)^\top$.  
Using the notation of empirical process, 
one can write 
\[
	M_{n,\tau}(\btheta)=\mathbb{P}_nm_{\btheta}\quad 
	\text{and}\quad  
	M_\tau(\btheta)=\mbox{P}m_{\btheta}, 
\] 
where $\mathbb{P}_n=n^{-1}\sum_{i=1}^n \delta_{\mathcal{X}_i}$ is the empirical measure, 
and $m_{\btheta}(\mathcal{X})=\rho_\tau\left(Y-\bxi^\top \bV(t)\right)
=\omega_{\tau}[Y-\bxi^\top \bV(t)]^2$ 
with the weights 
$$
      \omega_\tau(\mathcal{X})=\left|\tau-I(Y-\bxi^\top \bV(t)\leq 0)\right|=
      \begin{cases} 
      (1-\tau), & Y-\bxi^\top \bV(t)\leq 0, \\
      \tau , &  Y-\bxi^\top \bV(t)> 0. 
   \end{cases}.
$$ 
Here, $\mathcal{X}$ is the observed data $(Y, X, \bZ)$.   

In Lemma A.1 in the Appendix, we show that 
$\sup_{\btheta\in \Theta}\left|M_{n,\tau}(\btheta)-M_\tau(\btheta)\right|$ converges to zero in probability, as $n$ goes to infinity. Furthermore, we establish the consistency of 
$\widehat{\btheta}$. 

\begin{theorem}\label{thm1}
	Under the regularity conditions  in the Appendix,   as $n\rightarrow \infty$,  
	we have that 
	$
	\widehat{\btheta}\mathop{\longrightarrow}\limits^P \btheta_0.
	$
\end{theorem}

We prove the asymptotic normality by using  Theorem~5.23 in  \citet{van2000asymptotic}, 
which establishes the asymptotic normality of M-estimators when the criterion function is Lipschitz continuous and its limiting function admits a second order Taylor expansion.  
To proceed,  define the matrix 
$\Sigma(\btheta)=\mbox{E}\dot{m}_{\btheta}\dot{m}_{\btheta}^\top$,  
where $\dot{m}_{\btheta}$ is 
\[
\dot{m}_{\btheta}=
  \begin{bmatrix}
   -2\omega_\tau\bV(t)\left\{Y-\bxi^\top \bV(t)\right\} & \\
   2\beta_2 \mbox{E}\left\{\omega_\tau\left[Y- \bxi^\top \bV(t)\right]\bigg | X\right\} I(X>t) & 
  \end{bmatrix}.
\]
Define the Hessian matrix of $M_\tau(\btheta)$ 

\scalebox{0.7}{
 \begin{minipage}{0.8\linewidth}
\begin{align*}
	H(\btheta) 
	&\equiv \frac{\partial^2}{\partial \btheta \partial \btheta^\top}M_\tau (\btheta)\\
	&= 
	  2\mbox{E}\left( \omega_\tau
	 \begin{bmatrix}
     \bV(t)\bV(t)^\top & -\beta_2I(X>t) \bV(t) + \left\{Y-\bxi^\top \bV(t)\right\} \bU(t)\\
     -\beta_2I(X>t) \bV(t)^\top + \left\{Y-\bxi^\top \bV(t)\right\} \bU(t)^\top & \beta_2^2I(X>t)
  \end{bmatrix}
  \right)\\
  &+
   2\mbox{E}\left(
	\begin{bmatrix}
     \bm{0}_{(p+3)\times (p+3)} & \bm{0}_{(p+3)\times 1}\\
     \bm{0}_{1\times (p+3)} & -\beta_2\mbox{E}\left\{\omega_\tau
     \left[Y-\bxi^\top\bV(t) \right]\bigg|X=t\right\}f_X(t)
  \end{bmatrix} 
  \right),
\end{align*}
\end{minipage}
}
where $\bU(t)=[0,0, I(X>t),\bm{0}_{p\times 1}]^\top$. 

\begin{theorem}\label{thm2}
	Under the regularity conditions in the Appendix,   
	$\sqrt{n}(\widehat{\btheta}-\btheta_0)$ 
	is asymptotically normally distributed with  mean zero and covariance matrix 
	$H(\btheta_0)^{-1}\Sigma(\btheta_0) H(\btheta_0)^{-1}$, 
	as $n\rightarrow \infty$. 	
\end{theorem}

It is worthwhile to emphasize that the regression coefficients  and threshold estimators $(\widehat{\bxi}^\top, \widehat{t})^\top$ are jointly asymptotically normal with $\sqrt{n}$ convergence rate, 
and have non-zero asymptotic covariance in our model setting.    
This is different from the model with a discontinuous threshold. 
In the latter situation, 
the estimators of the regression coefficients  $\widehat{\bxi}$ are still $\sqrt{n}$-consistent, 
but the threshold estimator $\widehat{t}$  is $n$-consistent with a non-standard asymptotic distribution. 
The $\sqrt{n}$-convergence rate of $\widehat{t}$ in our model is due to the continuity of $M_{n,\tau}(\btheta)$ at $t$. 

The asymptotic variance-covariance matrix can be estimated by 
$\widehat{H}_n(\widehat{\btheta})^{-1}\widehat{\Sigma}(\widehat{\btheta})\widehat{H}_n(\widehat{\btheta})^{-1}$, 
where 
$\widehat{\Sigma}_n(\widehat{\btheta})=n^{-1}\sum_{i=1}^n \widehat{G}_n(\widehat{\btheta})  \widehat{G}_n(\widehat{\btheta})^\top$,  
and 

\scalebox{0.7}{
 \begin{minipage}{0.8\linewidth}
\begin{align*}
	\widehat{G}_n(\widehat{\btheta})
	&=
	 \begin{bmatrix}
   -2 \widehat{\omega}_{\tau,i} \bV_i(t)\left\{Y_i-\widehat{\bxi}^\top \bV_i(t)\right\} & \\
   2\widehat{\beta}_2  \widehat{\omega}_{\tau,i}\left\{Y_i-\widehat{\bxi}^\top \bV_i(t)\right\} I(X_i>t) & 
  \end{bmatrix},\\
	\widehat{H}_n(\widehat{\btheta})
	&=
	  \frac{2}{n}\sum_{i=1}^n \widehat{\omega}_{\tau,i}
	\begin{bmatrix}
     \bV_i(t)\bV_i(t)^\top & -\widehat{\beta}_2I(X_i>t) \bV_i(t) +
     \left\{Y_i-\bxi^\top \bV_i(t)\right\} \bU_i(t)\\
     -\widehat{\beta}_2I(X_i>t) \bV_i(t)^\top + \left\{Y_i-\bxi^\top \bV_i(t)\right\} \bU_i(t)^\top & \widehat{\beta}_2^2I(X_i>t)
  \end{bmatrix}\\
  &+
   \frac{2}{n}\sum_{i=1}^n 
	\begin{bmatrix}
     \bm{0}_{(p+3)\times (p+3)} & \bm{0}_{(p+3)\times 1}\\
     \bm{0}_{1\times (p+3)} & -\widehat{\beta}_2\widehat{\omega}_{\tau,i}\left\{Y_i-\bxi^\top \bV_i(t)\right\}\widehat{f}_X(t)
  \end{bmatrix}. 
\end{align*}
\end{minipage}
}
Here,  $ \widehat{\omega}_{\tau,i}=|\tau-I(Y_i-\widehat{\bxi}^\top \bV_i(t))|$, 
and $\widehat{f}_X(x)=(nh)^{-1}\sum_{i=1}^n K(\frac{X_i-x}{h})$ is the kernel estimator for the density $f_X(x)$ of $X$,  
and $K(\cdot)$ is a kernel function with a bandwidth $h>0$.  
In practice, we use the Epanechnikov kernel $K(u)=3/4 (1-u^2) I(|u|\leq 1)$ and 
obtain the optimal bandwidth by the Silverman's rule of thumb \citep{silverman1986density},  
$h=1.06\widehat{\sigma}n^{-1/5}$,  
where $\widehat{\sigma}$ is the standard deviation of $X$.

\subsection{Testing for structural change at a given expectile}
\label{ss:test}

An important question before fitting model \eqref{mod1} is whether there exists a threshold at a pre-specified  expectile.  
If a threshold does not exist,  then $t$ is unidentifiable and the estimation procedure in the last section is ill-conditioned.  
To test the existence of a threshold, we test null ($H_0$) and alternative ($H_1$) hypotheses 
\[
	H_0: \beta_2=0 \quad \text{for any $t \in \mathcal{T}$  v.s.} \quad  
	H_1: \beta_2\neq 0 \quad \text{for some $t \in \mathcal{T}$}, 
\]
where $\mathcal{T}$ is the range set of all $t$'s.

Tests for structural changes have been developed in conditional mean regression 
\citep{andrews1993tests, bai1996testing, hansen1996inference, hansen2015regression},  
quantile regression \citep{qu2008testing, li2011bent}, 
transformation models \citep{kosorok2007inference}, 
time series models \citep{chan1993consistency, cho2007testing},   and among others.  
To construct our test statistic, we take an approach in spirit similar to the test for structural changes in quantile regression in \citet{qu2008testing}. 
This test is constructed by sequentially evaluating the subgradients of the objective function under $H_0$ for a subsample, in a fashion similar to the CUSUM statistic. 
An advantage of this test is that it only requires fitting the model under the null hypothesis.  
Thus, it is computationally more efficient than the likelihood-ratio type tests, 
such as the  sup-likelihood-ratio-type test for testing threshold effects in regression models in  \citet{lee2011testing},  
which requires fitting the models under both null and alternative hypotheses.  

To proceed,  we define the following statistic, 
\[
	R_n(t)=\frac{1}{\sqrt{n}}\sum_{i=1}^n 
	\left|\tau-I(Y_i\leq \widehat{\balpha}^\top \bW_i)\right|(Y_i-\widehat{\balpha}^\top \bW_i)
	(X_i-t)I(X_i\leq t),
\]
where $\bW_i=(1, X_i, \bZ_i^\top)^\top$,  
and $\widehat{\balpha}$ is the estimator of coefficients $\balpha=(\beta_0, \beta_1, \bgamma^\top)^\top$ 
under the null hypothesis $H_0$, that is, 
\begin{align*}
	\widehat{\balpha}=\arg\min_{\balpha}\frac{1}{\sqrt{n}} \sum_{i=1}^n 
	\left|\tau-I(Y_i\leq \balpha^\top \bW_i)\right|(Y_i- \balpha^\top \bW_i)^2.
\end{align*}
An intuitive interpretation for $R_n(t)$ is given as follows.
If there is not a threshold, $\widehat{\balpha}$ is a good estimate of its population value, 
and hence,  the estimated residual $e_i=Y_i-\widehat{\balpha}^T \mathbf{W}_i$ would show a random pattern against $X_i$, leading to a small $R_n(t)$. 
On the other hand, if there exists a threshold, the estimate $\widehat{\balpha}$ would differ significantly from the true value,  and the estimated residuals would depart from zero in a systematic fashion related to $X_i$, resulting in a large absolute value of $R_n(t)$.  
Because the location of the threshold is unknown,  
we need search through all the possible locations.  
Therefore, we propose the test statistic 
 \[
 	T_n =\sup_{t\in \mathcal{T}}\left|R_n(t)\right|. 
 \]
This statistic can be viewed as a weighted CUSUM statistic based on the estimated residuals under the null hypothesis. 
Intuitively, it is plausible to reject $H_0$ when $T_n$ is too large. 
This intuition will be formally verified by Theorem \ref{thm3}. 
It implies that  $R_n(t)$ converges to a Gaussian process with  mean zero,  
and the size of such a process can be used to test for a threshold effect.  

In order to derive the large-sample inference for $T_n$,  
we consider the local alternative model, 
\begin{align}\label{mod2}
	Y_i = \beta_0 + \beta_1 X_i + n^{-1/2}\beta_2 (X_i-t)_+ + \bgamma^\top \bZ_i + e_i, 
\end{align}
where $t$ is the location of threshold,   
$\beta_2\neq 0$, 
and the $\tau$-expectile of $e_i$ is zero. 
We first introduce some notations
\begin{align*}
	\widehat{S}_{wn}(\widehat{\balpha}) &= 
	n^{-1}\sum_{i=1}^n \left|\tau-I(Y_i\leq \widehat{\balpha}^\top \bW_i)\right| \bW_i \bW_i^\top, \\
	S_w(\balpha) &=
	\mbox{E}\left[\left|\tau-I(Y\leq\balpha^\top \bW_1)\right| \bW_1 \bW_1^\top\right], \\
	\widehat{S}_{1n}(\widehat{\balpha},t) &=
	n^{-1}\sum_{i=1}^n \left|\tau-I(Y_i\leq \widehat{\balpha}^\top \bW_i)\right| \bW_i (X_i-t)I(X_i\leq t),  \\	
	S_1(\balpha, t) &= 
	\mbox{E}\left[\left|\tau-I(Y\leq \balpha^\top \bW_1)\right| \bW_1
	(X-t)I(X\leq t)\right], \\
	\widehat{S}_{2n}(\widehat{\balpha}, t) &= 
	n^{-1}\sum_{i=1}^n \left|\tau-I(Y_i\leq \widehat{\balpha}^\top \bW_i)\right| \bW_i \beta_2(X_i-t)I(X_i\geq t),  \\	
	S_2(\balpha, t) &= 
	\mbox{E}\left[\left|\tau-I(Y\leq \balpha^\top \bW_1)\right| \bW_1 
	\beta_2(X-t)I(X\geq t)\right], 
\end{align*}
and $q(t)= S_1(\balpha, t)^\top S_w(\balpha)^{-1}S_2(\balpha, t)$. 
\begin{theorem}\label{thm3}
	Under the regularity conditions in the Appendix,  for the local alternative model \eqref{mod2}, 
	$R_n(t)$ has the asymptotic representation
	\begin{align}\label{asym}
		R_n( t) 
		&=\frac{1}{\sqrt{n}}\sum_{i=1}^n e_i
		\left|\tau-I(Y_i-\balpha^\top \bW_i\leq 0)\right|
		\left[(X_i-t)I(X_i\leq t)- S_1(\balpha, t)^\top S_w(\balpha)^{-1}\bW_i\right] \\\nonumber
		&-q(t)+o_P(1). 
	\end{align}
	Furthermore, $T_n$ converges weakly to the process $\sup_t|R(t)-q(t)|$, 
	where $R(t)$ is the Gaussian process with mean zero and covariance function 
	\begin{eqnarray*}
    &&\hbox{E}
  	\bigg[
  	e_1^2
  \left|\tau-I(Y_1-\balpha^\top \bW_1\leq 0)\right|
  \left\{
  (X_1-t_1)I(X_1\leq t_1)  - S_1(\balpha, t_1)^TS_w(\balpha)^{-1}\bW_1
  \right\}\\
  &&\times 
  \left\{
  (X_1-t_2)I(X_1\leq t_2)  - S_1(\balpha, t_2)^TS_w(\balpha)^{-1}\bW_1
  \right\}
  	\bigg]. 
	\end{eqnarray*}
	
\end{theorem}

\begin{corollary}\label{coro1}
	Under the regularity conditions in the Appendix,  for the local alternative model,   
	$Y_i = \beta_0 + \beta_1 X_i + n^{-1/2}a_n\beta_2 (X_i-t)_+ + \bgamma^\top \bZ + e_i$, 
	for any increasing sequence $a_n$ goes to infinite, we have that $\mathop{\lim}\limits_{n\rightarrow \infty}P(|T_n|\geq t)=1$ for any $t>0$. 	
\end{corollary}

Because the limiting null distribution of $T_n$ is nonstandard, we resort to the Gaussian multiplier method  \citep{van2000asymptotic} to calculate the critical values, 
based on the asymptotic representation \eqref{asym}.  
The procedure is described in Algorithm~1. 
In the Appendix, we prove the following result,  which implies the validity of the bootstrap resampling scheme.   
\begin{theorem}\label{thm3}
	Under both the null and the local alternative hypotheses, $R_n^*(\tau)$ (defined in Algorithm~1) converges to the Gaussian process $R(t)$ as $n\rightarrow \infty$. 
\end{theorem}

We summarize the computing procedure as follows. 

\fbox{
\begin{minipage}{\dimexpr\textwidth-2\fboxsep-2\fboxrule\relax} 
	\parbox{1\textwidth}{%
		\textbf{Algorithm~1}:
		\begin{center} 
			\begin{description}
				\item[1]  
				Generate iid $\{v_1,\cdots, v_n\}$  from    $N(0, 1)$. 
				
				\item[2]  
				Calculate the test statistic $T_n^*(t)=\sup_{t\in\mathcal{T}} 
				\left|R_n^*(t)\right|$, 
				where 
				\begin{align*}
				R_n^*(t)
				&= \frac{1}{\sqrt{n}}\sum_{i=1}^{n} v_i
				\widehat{e}_i
				\left|\tau-I(\widehat{e}_i\leq 0)\right|\\
				& \times
				\left[
				(X_i-t)I(X_i\leq t)- 
				\widehat{S}_{1n}(\widehat{\balpha}, t)^\top 
				\widehat{S}_{wn}(\widehat{\balpha})^{-1}\bW_i
				\right],
				\end{align*}		
				with the estimated residuals 
				$\widehat{e}_i=Y_i-\widehat{\balpha}^\top \bW_i$ under the null hypothesis.  
				
				\item[3] 
				Repeat Steps~1--2 with $\hbox{NB}$ times to obtain $T_{n}^{*(1)},
				\cdots, T_n^{*(\hbox{NB})}$. 
				Calculate the p-value as  
				$\widehat{p}_n =\hbox{NB}^{-1}\mathop{\sum}\limits_{j=1}^{\hbox{NB}}
				 I\{T_n^{*(j)}\geq T_n\}$. 	
			\end{description}
		\end{center}
	}%
\end{minipage}
}%

\section{Simulation Studies and Applications}
\label{s:simulation}
\subsection{Simulation studies}

In this section, we conduct simulation studies for assessing the finite sample performance of the proposed method. We consider the following two scenarios: 
\begin{itemize}
  \item[(i)] 
	 Independent and identically distributed~(IID):
	 $
	Y=\beta_0+\beta_1X+\beta_2(X-t)_++\gamma Z+e, 
	$
		
  \item[(ii)]
	Heteroscedasticity: 
	$
	Y=\beta_0+\beta_1X+\beta_2(X-t)_++\gamma Z+(1+0.2Z)e, 
	$
\end{itemize}
where  $x$ is generated from a uniform distribution $U(-2, 4)$, 
$z$ is generated from a normal distribution $N(1, 0.5^2)$,  
and the parameters are $(\beta_0, \beta_1, \beta_2, \gamma, t)^\top=(1, 3, -2,  1, 1.5)^\top$. 
For each scenario, we consider three error cases: (1)  $e\sim N(0,1)$,  
(2) $e\sim t_4$,  
and 
(3) a mixture distribution $e\sim 0.9N(0,1)+0.1t_4$,  
where $t_4$ is the $t$-distribution with four degrees of freedom.  
For each case, we conduct 1000 repetitions with sample sizes $n=200$ and $400$. 

As shown in Tables~\ref{tab:sim1}---\ref{tab:sim2},  
for both the IID and the heteroscedastic scenarios,  
all the biases are small, indicating the proposed estimator is asymptotically consistent.  
Moreover, the average estimated standard errors are close to the empirical standard errors. 
The coverage probabilities of the regression parameters $(\beta_0, \beta_1, \beta_2, \gamma)$ are close to the nominal level $95\%$. 
Though some coverage probabilities of the threshold $t$ are below $90\%$ when $n=200$, 
they improve as the sample size increases to $n=400$.  
The performance is similar in all the three error distributions.
In summary, the proposed estimate has a good finite sample performance.

We also conduct simulation studies to evaluate the type I error and the power of the testing procedure.  
The simulation models are similar to the above,  
with threshold effects at $\beta_2= -2, -1, -0.5, 0, 0.5, 1, 2$. 
The number of bootstrap times is set as $1,000$ and the nominal significance level is $5\%$. 
The results are shown in Table~\ref{tab:test}. 
For all scenarios, 
the tests have type I errors close to the nominal level and have reasonable power, 
which indicates that the proposed test is valid for testing the existence of a threshold.

\begin{table}
   \caption{Performance of the proposed estimator based on 1,000 simulated samples of 
   $n=200$ and $400$ observations, for the three error distributions in the  IID case. }\label{tab:sim1}
\begin{center}
\scalebox{0.7}{
\begin{tabular}{crcrrrrrcrrrrr}
  \hline\hline
  \multicolumn{3}{c}{}& \multicolumn{4}{c}{$n=200$} & &  \multicolumn{5}{c}{$n=400$}\\
				\cline{4-8}\noalign{}  \cline{10-14}\noalign{}
Error & $\tau$ &  & $\beta_0$ & $\beta_1$ & $\beta_2$ & $\gamma$ & $t$ & &
 $\beta_0$ & $\beta_1$ &$\beta_2$  & $\gamma$ & $t$ \\ 
  \hline
1 & & True & 1.000 & 3.000 & -2.000 & 1.000 & 1.500 && 1.000 & 3.000 & -2.000 & 1.000 & 1.500 \\ 
   &0.3& Bias & 0.012 & 0.006 & -0.013 & -0.002 & -0.008 && 0.003 & 0.003 & -0.006 & 0.001 & -0.002 \\ 
   && SD & 0.172 & 0.101 & 0.187 & 0.140 & 0.176 && 0.135 & 0.070 & 0.131 & 0.107 & 0.121 \\ 
   && ESE & 0.175 & 0.094 & 0.181 & 0.145 & 0.146 && 0.124 & 0.067 & 0.129 & 0.103 & 0.104 \\ 
   && CP & 0.944 & 0.926 & 0.947 & 0.950 & 0.877 && 0.923 & 0.934 & 0.951 & 0.939 & 0.915 \\ 
   & 0.5& Bias & 0.008 & 0.005 & -0.012 & -0.002 &-0.006 && -0.000 & 0.002 & -0.007 & 0.002 & -0.001 \\ 
   && SD & 0.168 & 0.098 & 0.180 & 0.136 & 0.170 && 0.132 & 0.067 & 0.126 & 0.105 & 0.116 \\ 
   && ESE & 0.170 & 0.091 & 0.177 & 0.140 & 0.142 && 0.120 & 0.065 & 0.125 & 0.100 & 0.101 \\ 
   && CP & 0.947 & 0.928 & 0.945 & 0.953 & 0.889 && 0.932 & 0.941 & 0.947 & 0.941 & 0.917 \\ 
   &0.8 & Bias & 0.000 & 0.006 & -0.015 & -0.002 & -0.007 && -0.004 & 0.003 & -0.010 & 0.002 & -0.001 \\ 
   && SD & 0.186 & 0.108 & 0.197 & 0.150 & 0.187 && 0.140 & 0.073 & 0.139 & 0.113 & 0.126 \\ 
   && ESE & 0.182 & 0.098 & 0.191 & 0.149 & 0.153 && 0.130 & 0.070 & 0.136 & 0.107 & 0.110 \\ 
   && CP & 0.942 & 0.927 & 0.935 & 0.946 & 0.880 && 0.927 & 0.939 & 0.944 & 0.934 & 0.900 \\ 
2 & 0.3 & Bias & 0.010 & 0.023 & -0.058 & -0.001 & -0.007 && 0.009 & 0.006 & -0.028 & -0.003 & -0.000 \\ 
   && SD & 0.279 & 0.148 & 0.302 & 0.225 & 0.275 && 0.187 & 0.111 & 0.200 & 0.151 & 0.190 \\ 
   && ESE & 0.254 & 0.137 & 0.269 & 0.212 & 0.209 && 0.183 & 0.098 & 0.193 & 0.152 & 0.153 \\ 
   && CP & 0.928 & 0.927 & 0.920 & 0.938 & 0.858 && 0.948 & 0.918 & 0.940 & 0.962 & 0.891 \\ 
   & 0.5 & Bias & 0.002 & 0.016 & -0.049 & 0.001 & 0.001 && 0.006 & 0.004 & -0.023 & -0.002 & 0.000 \\ 
   && SD & 0.257 & 0.135 & 0.278 & 0.206 & 0.261 && 0.175 & 0.101 & 0.184 & 0.140 & 0.172 \\ 
   && ESE & 0.237 & 0.128 & 0.256 & 0.197 & 0.199 && 0.170 & 0.091 & 0.181 & 0.140 & 0.144 \\ 
   && CP & 0.930 & 0.926 & 0.937 & 0.947 & 0.865 && 0.957 & 0.925 & 0.937 & 0.956 & 0.901 \\ 
   &0.8& Bias & 0.002 & 0.017 & -0.072 & 0.003 & 0.003 && 0.007 & 0.010 & -0.038 & -0.001 & -0.001 \\ 
   && SD & 0.326 & 0.176 & 0.362 & 0.256 & 0.342 && 0.228 & 0.132 & 0.244 & 0.180 & 0.233 \\ 
   && ESE & 0.453 & 0.335 & 0.544 & 0.240 & 0.455 && 0.213 & 0.119 & 0.241 & 0.171 & 0.186 \\ 
   &&CP & 0.924 & 0.935 & 0.933 & 0.939 & 0.856 && 0.948 & 0.915 & 0.941 & 0.939 & 0.893 \\ 
3 & 0.3& Bias & 0.015 & 0.001 & -0.013 & -0.004 & -0.002 && 0.011 & 0.003 & -0.012 & -0.006 & -0.000 \\ 
   && SD & 0.197 & 0.108 & 0.237 & 0.162 & 0.201 && 0.134 & 0.073 & 0.145 & 0.108 & 0.125 \\ 
   && ESE & 0.183 & 0.098 & 0.196 & 0.152 & 0.155 && 0.130 & 0.070 & 0.136 & 0.108 & 0.109 \\ 
   && CP & 0.930 & 0.923 & 0.927 & 0.928 & 0.881 && 0.940 & 0.938 & 0.940 & 0.954 & 0.910 \\ 
   & 0.5 & Bias & 0.010 & -0.000 & -0.012 & -0.004 & -0.000 && 0.009 & 0.003 & -0.010 & -0.005 & -0.001 \\ 
  && SD & 0.189 & 0.102 & 0.213 & 0.156 & 0.191 && 0.127 & 0.070 & 0.138 & 0.102 & 0.120 \\ 
  && ESE & 0.178 & 0.095 & 0.188 & 0.146 & 0.150 && 0.125 & 0.068 & 0.131 & 0.104 & 0.106 \\ 
  && CP & 0.935 & 0.932 & 0.926 & 0.932 & 0.890 && 0.948 & 0.940 & 0.942 & 0.957 & 0.918 \\ 
  & 0.8 & Bias & 0.005 & 0.002 & -0.022 & -0.005 & -0.000 && 0.009 & 0.003 & -0.009 & -0.005 & -0.006 \\ 
  && SD & 0.208 & 0.116 & 0.234 & 0.171 & 0.217 && 0.140 & 0.081 & 0.157 & 0.114 & 0.138 \\ 
  && ESE & 0.195 & 0.106 & 0.219 & 0.160 & 0.170 && 0.141 & 0.077 & 0.148 & 0.115 & 0.119 \\ 
  && CP & 0.930 & 0.929 & 0.929 & 0.929 & 0.887 && 0.944 & 0.945 & 0.936 & 0.956 & 0.918 \\ 
   \hline
\end{tabular}}
\end{center}
{\footnotesize Bias: the empirical bias;  	SD: the empirical standard error;  
ESE: the average estimated standard error;  CP: $95\%$ coverage probability.	}
\end{table}

\begin{table}
   \caption{Performance of the proposed estimator based on 1,000 simulated samples of 
   $n=200$ and $400$ observations, for the three error distributions in the heteroscedastic case. }\label{tab:sim2}
\begin{center}
\scalebox{0.7}{
\begin{tabular}{crcrrrrrcrrrrr}
  \hline\hline
  \multicolumn{3}{c}{}& \multicolumn{4}{c}{$n=200$} & &  \multicolumn{5}{c}{$n=400$}\\
				\cline{4-8}\noalign{}  \cline{10-14}\noalign{}
Error & $\tau$ &  & $\beta_0$ & $\beta_1$ & $\beta_2$ & $\gamma$ & $t$ & &
 $\beta_0$ & $\beta_1$ &$\beta_2$  & $\gamma$ & $t$ \\ 
  \hline
&& True & 1.000 & 3.000 & -2.000 & 1.000 & 1.500 && 1.000 & 3.000 & -2.000 & 1.000 & 1.500 \\ 
1 & 0.3 & Bias & 0.013 & 0.008 & -0.024 & 0.000 & -0.004 && 0.004 & 0.005 & -0.010 & 0.001 & -0.004 \\ 
& & SD & 0.199 & 0.125 & 0.227 & 0.170 & 0.219 && 0.154 & 0.085 & 0.157 & 0.129 & 0.149 \\ 
& & ESE & 0.200 & 0.113 & 0.219 & 0.175 & 0.175 && 0.141 & 0.080 & 0.156 & 0.124 & 0.125 \\ 
& &CP & 0.958 & 0.921 & 0.946 & 0.948 & 0.866 && 0.928 & 0.927 & 0.947 & 0.935 & 0.905 \\ 
& 0.5 & Bias & 0.010 & 0.007 & -0.022 & -0.004 & -0.003 && 0.002 & 0.004 & -0.011 & 0.001 & -0.001 \\ 
& & SD & 0.194 & 0.119 & 0.218 & 0.165 & 0.209 && 0.150 & 0.082 & 0.152 & 0.126 & 0.144 \\ 
& & ESE & 0.195 & 0.110 & 0.214 & 0.170 & 0.171 && 0.137 & 0.078 & 0.151 & 0.121 & 0.122 \\ 
& & CP & 0.952 & 0.930 & 0.948 & 0.949 & 0.880 && 0.931 & 0.940 & 0.947 & 0.941 & 0.903 \\ 
&0.8& Bias & 0.008 & 0.009 & -0.027 & -0.009 & -0.007 && 0.000 & 0.004 & -0.016 & -0.002 & -0.000 \\ 
&& SD & 0.214 & 0.130 & 0.241 & 0.182 & 0.235 && 0.160 & 0.090 & 0.168 & 0.136 & 0.160 \\ 
& & ESE & 0.209 & 0.119 & 0.232 & 0.181 & 0.185 && 0.148 & 0.084 & 0.165 & 0.130 & 0.132 \\ 
&& CP & 0.944 & 0.928 & 0.934 & 0.945 & 0.881 && 0.934 & 0.936 & 0.947 & 0.936 & 0.894 \\ 
2& 0.3 & Bias & 0.015 & 0.035 & -0.086 & 0.002 & -0.015 && 0.010 & 0.007 & -0.042 & -0.003 & 0.006 \\ 
&  & SD & 0.323 & 0.183 & 0.366 & 0.272 & 0.346 && 0.215 & 0.134 & 0.240 & 0.183 & 0.230 \\ 
&  & ESE & 0.293 & 0.168 & 0.328 & 0.256 & 0.251 && 0.209 & 0.118 & 0.234 & 0.184 & 0.184 \\ 
 & & CP & 0.927 & 0.923 & 0.918 & 0.937 & 0.839 && 0.951 & 0.912 & 0.945 & 0.959 & 0.881 \\ 
&0.5& Bias & 0.008 & 0.024 & -0.074 & 0.001 & -0.002 && 0.009 & 0.006 & -0.034 & -0.004 & 0.003 \\ 
&& SD & 0.297 & 0.166 & 0.338 & 0.249 & 0.323 && 0.200 & 0.121 & 0.221 & 0.169 & 0.212 \\ 
& & ESE & 0.274 & 0.157 & 0.316 & 0.239 & 0.241 && 0.194 & 0.110 & 0.218 & 0.170 & 0.172 \\ 
&& CP & 0.937 & 0.924 & 0.939 & 0.945 & 0.846 && 0.957 & 0.927 & 0.938 & 0.955 & 0.897 \\ 
&0.8& Bias & 0.020 & 0.029 & -0.105 & -0.004 & -0.004 && 0.018 & 0.016 & -0.057 & -0.007 & -0.002 \\ 
&& SD & 0.374 & 0.216 & 0.437 & 0.310 & 0.410 && 0.264 & 0.161 & 0.297 & 0.216 & 0.289 \\ 
& & ESE & 0.364 & 0.226 & 0.499 & 0.287 & 0.346 && 0.243 & 0.143 & 0.307 & 0.206 & 0.230 \\ 
& & CP & 0.924 & 0.936 & 0.931 & 0.935 & 0.851 && 0.944 & 0.915 & 0.945 & 0.937 & 0.884 \\ 
3&0.3 & Bias & 0.015 & 0.002 & -0.025 & -0.001 & 0.001 && 0.013 & 0.005 & -0.018 & -0.006 & 0.001 \\ 
& & SD & 0.226 & 0.131 & 0.285 & 0.197 & 0.242 && 0.153 & 0.089 & 0.176 & 0.129 & 0.156 \\ 
& & ESE & 0.210 & 0.118 & 0.239 & 0.183 & 0.186 && 0.148 & 0.084 & 0.164 & 0.131 & 0.132 \\ 
 && CP & 0.937 & 0.920 & 0.923 & 0.930 & 0.872 && 0.941 & 0.937 & 0.937 & 0.950 & 0.897 \\ 
&0.5& Bias & 0.012 & 0.001 & -0.022 & -0.005 & 0.002 && 0.012 & 0.005 & -0.015 & -0.007 & -0.003 \\ 
& & SD & 0.215 & 0.123 & 0.258 & 0.188 & 0.229 && 0.144 & 0.086 & 0.166 & 0.122 & 0.148 \\ 
&& ESE & 0.203 & 0.115 & 0.228 & 0.177 & 0.181 && 0.143 & 0.082 & 0.158 & 0.126 & 0.127 \\ 
& & CP & 0.931 & 0.931 & 0.929 & 0.931 & 0.875 && 0.949 & 0.943 & 0.944 & 0.957 & 0.919 \\ 
&0.8& Bias & 0.013 & 0.004 & -0.035 & -0.013 & 0.002 && 0.015 & 0.005 & -0.015 & -0.010 & -0.007 \\ 
& &SD & 0.239 & 0.142 & 0.282 & 0.206 & 0.264 && 0.159 & 0.099 & 0.190 & 0.138 & 0.169 \\ 
&& ESE & 0.224 & 0.129 & 0.329 & 0.193 & 0.225 && 0.162 & 0.094 & 0.180 & 0.140 & 0.144 \\ 
&& CP & 0.930 & 0.926 & 0.927 & 0.927 & 0.877 && 0.946 & 0.937 & 0.935 & 0.949 & 0.913 \\ 
  \hline
\end{tabular}
}
\end{center}
{\footnotesize Bias: the empirical bias;  	SD: the empirical standard error;  
ESE: the average estimated standard error;  CP: $95\%$ coverage probability.	}
\end{table}

\begin{table}
\caption{Power analysis for IID and heteroscedastic models with three different errors, 
based on 1,000 simulated samples of  $n=200$ observations. }\label{tab:test}
\centering 
\scalebox{0.9}{
\begin{tabular}{cccccccccc}
  \hline\hline
Model & Error & $\tau$ &  \multicolumn{7}{c}{$\beta_2$} \\ 
\cline{4-10}\noalign{} 
&  &  & -2 & -1 & -0.5 & 0& 0.5 & 1 & 2 \\ 
IID &    1 & 0.3 & 1.000 & 1.000 & 0.770 & 0.049 & 0.766 & 1.000 & 1.000 \\ 
 &  & 0.5 & 1.000 & 1.000 & 0.801 & 0.052 & 0.788 & 1.000 & 1.000 \\ 
 &  & 0.8 & 1.000 & 0.999 & 0.745 & 0.048 & 0.712 & 1.000 & 1.000 \\ 
 &    2 & 0.3 & 0.998 & 0.925 & 0.442 & 0.065 & 0.478 & 0.935 & 0.999 \\ 
 &  & 0.5 & 1.000 & 0.974 & 0.492 & 0.063 & 0.503 & 0.960 & 1.000 \\ 
 &  & 0.8 & 0.994 & 0.860 & 0.400 & 0.067 & 0.377 & 0.873 & 0.993 \\ 
 &    3 & 0.3 & 1.000 & 0.992 & 0.710 & 0.035 & 0.729 & 0.998 & 1.000 \\ 
 &  & 0.5 & 1.000 & 0.996 & 0.738 & 0.041 & 0.769 & 0.999 & 1.000 \\ 
 &  & 0.8 & 1.000 & 0.990 & 0.682 & 0.039 & 0.662 & 0.990 & 0.998 \\ 
heteroscedastic &    
1 & 0.3 & 1.000 & 0.990 & 0.609 & 0.051 & 0.610 & 0.994 & 1.000 \\ 
 &  & 0.5 & 1.000 & 0.995 & 0.644 & 0.054 & 0.624 & 0.998 & 1.000 \\ 
 &  & 0.8 & 1.000 & 0.986 & 0.586 & 0.052 & 0.555 & 0.993 & 1.000 \\ 
 &    2 & 0.3 & 0.998 & 0.838 & 0.326 & 0.065 & 0.345 & 0.851 & 0.996 \\ 
 &  & 0.5 & 1.000 & 0.884 & 0.381 & 0.064 & 0.384 & 0.902 & 1.000 \\ 
 &  & 0.8 & 0.990 & 0.757 & 0.307 & 0.067 & 0.282 & 0.755 & 0.986 \\ 
 &    3 & 0.3 & 0.999 & 0.982 & 0.538 & 0.040 & 0.596 & 0.988 & 1.000 \\ 
 &  & 0.5 & 1.000 & 0.984 & 0.593 & 0.040 & 0.598 & 0.993 & 1.000 \\ 
 &  & 0.8 & 0.999 & 0.967 & 0.548 & 0.037 & 0.521 & 0.961 & 0.997 \\ 
   \hline
\end{tabular}
}
\end{table}

\subsection{Applications}
\subsubsection{Fourth Dutch growth data}

\begin{figure}
\centering
\subfloat[][Scatter plot of height (in cm) against age (in years) of Dutch boys .]{
\includegraphics[height=3in, width = 0.9\textwidth]{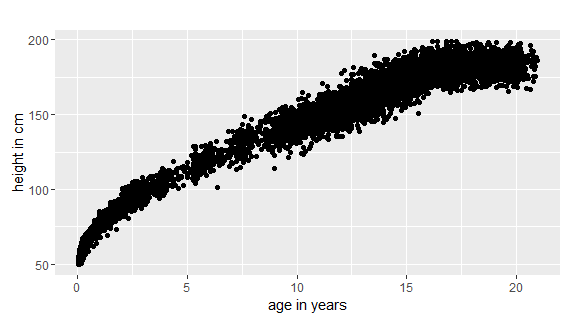}
\label{fig:scatter}}
\vspace{5mm}
\subfloat[][Fitted  expectile curves for the data with transformed age.]{
\includegraphics[height=3in, width = 0.9\textwidth]{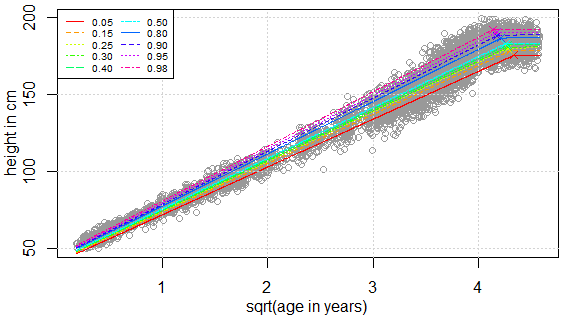}
\label{fig:fit}}
\caption{Analysis of Dutch boys data from the Fourth Dutch Growth Study. }
\label{fig:dutch}
\end{figure}

We first apply our method to the Fourth Dutch Growth data, which was collected by the Fourth Dutch Growth study \citep{van2007worm}  and is available in the \textbf{R} package \textit{expectreg}. 
This dataset has  the height, weight and head circumference 
of Dutch children between ages 0 and 21 years \citep{van2001worm}.    
A primary interest of this study concerns the relation between age and height.  
The scatter plot  (Figure~\ref{fig:scatter}) shows the relationship between age and height  for a  subset of $6,848$ boys. 
Clearly, there is a nonlinear trend between height and age (Figure~\ref{fig:scatter}), 
with a steep curvature  before age three due to rapid growth in early childhood,  
and a bent in the late teens due to reaching the full adult height. 
This dataset has been analyzed by  \citet{schnabel2009optimal}.  
In their analysis, they took a square root transformation on age.  
While this transformation effectively removes the curvature at early childhood, 
the nonlinearity in the late teens still exists (Figure~\ref{fig:fit}). 
Then they fitted the transformed data using smoothed expectile regression,  
by combining the least asymmetrically weighted squares with the P-splines.   
Though the smoothed expectile curves fit the data well,  
they do not provide any information on the location of the threshold, 
i.e., the age to stop growing.  

Here,  we fit the continuous threshold expectile model to the square root transformed data 
$(X_i, Y_i), i=1, \ldots, 6,848$  and estimate the location of threshold.  
Specifically, 
\begin{align}\label{mod:dutch}
	\nu_\tau(Y_i|X_i,\bZ) = \beta_0 + \beta_1 X_i + \beta_2 (X_i-t)_+ , 
\end{align}
where $Y_i$ is the height of the $i$th boy, $X_i$ is  the square root of  his age, 
and $\theta_\tau=(\beta_0, \beta_1, \beta_2)$ are the unknown parameters of interest, 
$t$ is the unknown location of threshold.  
We fit the model with
 $\tau=$ 0.05, 0.15, 0.25, 0.30, 0.40, 0.50, 0.80, 0.90, 0.95, 0.98. 

For all the expectile levels we fit, 
the p-values from our threshold effect test are nearly 0, 
indicating a highly significant continuous threshold pattern. 
The regression results for different expectile levels are reported in Table~4.  
The estimated coefficients show that the height first increases rapidly with age 
(roughly 31--35 cm per square root of age),  
and then the growth is very limited or nearly stops after about age 17--18.  
The estimated thresholds illustrate a general trend that shorter boys seem to stop growth later than taller boys, with a $95\%$ confidence interval (CI), [18.39, 19.24] 
at the expectile level $\tau=0.05$, 
and  [16.76, 17.40] at $\tau=0.98$. 
Figure~\ref{fig:fit} confirms these results. 

\begin{table}
\caption{The estimated parameters and their standard errors (listed in parentheses)  for Dutch boys data. The p-values are from the test for a threshold effect. }\label{tab:dutch}
\begin{minipage}{\textwidth}
\centering
\begin{tabular}{cccccc}
  \hline\hline
 $\tau$ &p-value &   $\beta_0$ & $\beta_1$ & $\beta_2$ & $t$ \\ 
  \hline
0.05& 0    & 40.875 & 31.019 & -30.268 & 4.337 \\ 
	  &  & (0.191) & (0.083) & (6.897) & (0.025) \\ 
0.15& 0  & 41.624 & 31.789 & -30.488 & 4.296 \\ 
       &   & (0.139) & (0.066) & (5.284) & (0.0239) \\ 
0.25&0  & 42.011 & 32.217 &-30.252 & 4.276 \\ 
	 &   & (0.123) & (0.061) & (4.263) & (0.021) \\ 
0.30&0   & 42.182 & 32.381 & -31.470 & 4.276 \\ 
       &  & (0.118) & (0.059) & (3.804) & (0.018) \\ 
0.40&0  & 42.491 & 32.670 & -31.727 & 4.276 \\ 
      & & (0.111) & (0.057) & (3.233) & (0.014) \\ 
0.50&0  & 42.751 & 32.956 & -32.305 & 4.255 \\ 
      &   & (0.107) & (0.057) & (3.625) & (0.018) \\ 
0.80&0  & 43.692 & 33.909 & -33.306 & 4.214 \\ 
      &  & (0.100) & (0.059) & (2.549) & (0.014) \\ 
0.90&0  & 44.235 & 34.456 & -31.624 & 4.173 \\ 
	 &  & (0.103) & (0.064) & (2.596) & (0.017) \\ 
0.95&0  & 44.780 & 34.887 & -34.831 & 4.173 \\ 
	 &  & (0.111) & (0.069) & (2.309) & (0.014) \\ 
0.98&0  & 45.444 & 35.427 & -34.072 & 4.133 \\ 
	 & & (0.129) & (0.080) & (2.582) & (0.020) \\ 
   \hline
\end{tabular}
\end{minipage}
\end{table}

\subsubsection{Baseball pitcher salary}

Our second example concerns the salaries of major league baseball (MLB) players for
the 1987 baseball season \citep{hoaglin1995critical}. 
The dataset has been analyzed by several groups in the ASA graphical session in 1989.  
Here we consider a subset with n=176 pitchers, 
which was analyzed in \citep{hettmansperger2011robust} using a rank-based regression.
This dataset is available in the \textbf{R} package \textit{rfit}.  
It consists of  the 1987 beginning salary and the number of years of experience for these pitchers.

\begin{figure}
\centering
\subfloat[][Fitted  expectile curves for the data.]{
\includegraphics[height=3in, width = \textwidth]{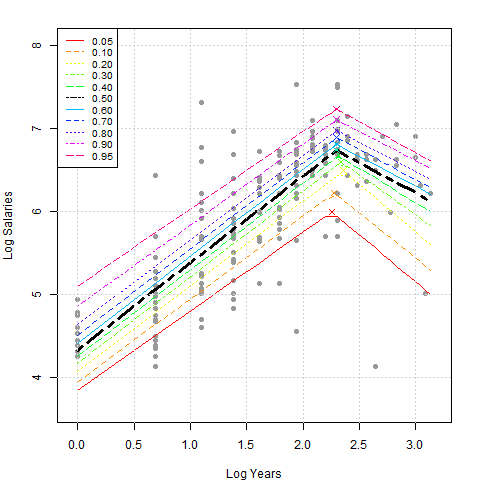}
\label{fig:bbsExp}}
\vspace{5mm}
\subfloat[][Fitted  quantile curves for the data.]{
\includegraphics[height=3in, width = \textwidth]{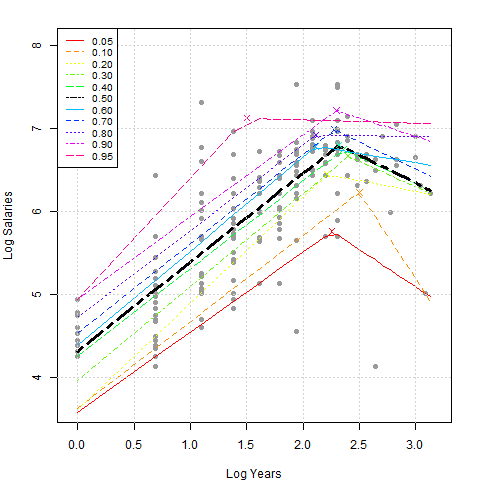}
\label{fig:bbsQnt}}
\caption{Analysis of baseball salaries data. }
\label{fig:bbsalaries}
\end{figure}

Visually,  the scatter plot (Figure~\ref{fig:bbsExp}) suggests that the salaries are first  positively correlated with the years of experience,  but then decline after about 9 years.  
This is somewhat unusual, because it is generally expected that salaries grow with the years of experience in players' early career and the status of free agent (i.e., the player whose initial 6 year contract expires).  
Although salaries do decrease after players pass their prime time,  
it would happen much later, for example,  \citet{haupert2012regime} estimated the decline for MLB players occurs after 22 years. 
In the analysis by the ASA graphical session in 1989,  
the model with the best predictive performance is a segmented mean regression model with a fixed threshold at 7 years,  where the threshold was chosen according to the length of the initial professional baseball contracts (6 years).  
It is of interest to formally test if the visually observed transition is significant 
and estimate  the onset of the decline from the data.  
Furthermore, the salaries show considerable heterogeneity at a given number of years of experience.  
Hence, a regression model based on the conditional distribution  of the response variable provides a more complete picture  than a mean regression model. 
Previous analyses only focused on the mean regression model \citep{hettmansperger2011robust, hoaglin1995critical},  
but not regression models for conditional distribution.  

Here we fit the data using the continuous threshold expectile regression, 
\begin{align}\label{mod:bbsalaries}
	\nu_\tau(Y_i|X_i,\bZ) = \beta_0 + \beta_1 X_i + \beta_2 (X_i-t)_+ , 
\end{align}
where $Y_i$ is the log (salary) of the $i$th pitcher, 
$X_i$ is log (years of experience),  
and $\theta_\tau=(\beta_0, \beta_1, \beta_2)$ are the unknown parameters of interest, 
$t$ is the unknown location of threshold,  
$\tau=$ 0.01, 0.02, 0.05, 0.1, 0.2, 0.3, 0.4, 0.5, 0.6, 0.7, 0.8, 0.9, 0.95, 0.98, 0.99. 

Our threshold test shows that the continuous threshold patterns are highly significant, with p-values less than 0.05 for all the expectile levels considered. 
Table~\ref{tab:bbsalaries} reports the estimated coefficients and their standard errors. 
The coefficients show that the salaries indeed decline  for pitchers with 9 or more years of experience (range: (8.61, 10.35)),  at all the expertile levels we fitted.  
Figure~\ref{fig:bbsalaries} confirms this conclusion.   

This raises two natural questions: why did the salaries decrease for more experienced pitchers? and why did the decrease occur at 9 years  for all salary levels?  
The history of the MLB shows that, 
in the time period of 1985 to 1987, the  MLB team owners colluded in an effort to decrease salaries for free agents after their initial contracts expired. 
Pitchers with 9 or more years of experience are all free agents. 
Their salary decrease is a reflection of owners trying to control salaries.  
The reason that the observed threshold (9 years) is later than the start of free agents (7 years), 
is that some pitchers have become free  agents before the collusion, 
thus they had more than 7 years of experience when the collusion occurred.  

As a comparison, 
we also fit the data with the bent-line quantile regression \citep{li2011bent}. 
Though the overall trend is similar to the continuous threshold expectile regression, 
it has more crossing between quantiles. 
This agrees with the observation of  \citet{schnabel2009optimal} and \citet{waltrup2015expectile}  
that expectile regression tends to have less crossing than quantile regression.

\begin{table}
\caption{The estimated parameters and their standard errors (listed in parentheses) for baseball salaries data. The p-value is testing for a threshold effect. }\label{tab:bbsalaries}
\centering
\scalebox{0.9}{
\begin{tabular}{cccccc}
  \hline
$\tau$ & p-value & $\beta_0$ & $\beta_1$ & $\beta_2$ & $t$ \\ 
  \hline
0.01 & 0.007 & 3.800 & 0.765 & -3.207 & 2.296 \\ 
&  & (0.224) & (0.242) & (1.481) & (0.201) \\ 
0.02 & 0.005 & 3.796 & 0.858 & -2.876 & 2.296 \\ 
&  & (0.195) & (0.206) & (1.379) & (0.111) \\ 
0.05 & 0.061 & 3.842 & 0.954 & -2.079 & 2.255 \\ 
&  & (0.134) & (0.130) & (0.309) & (0.139) \\ 
0.10 & 0.022 & 3.936 & 1.005 & -2.086 & 2.276 \\ 
&  & (0.134) & (0.117) & (0.488) & (0.141) \\ 
0.20 & 0.009 & 4.073 & 1.026 & -2.172 & 2.357 \\ 
&  & (0.104) & (0.078) & (0.591) & (0.136) \\ 
0.30 & 0.004 & 4.166 & 1.040 & -2.005 & 2.337 \\ 
&  & (0.095) & (0.066) & (0.586) & (0.122) \\ 
0.40 & 0.003 & 4.253 & 1.045 & -1.871 & 2.316 \\ 
&  & (0.087) & (0.057) & (0.530) & (0.116) \\ 
0.50 & 0.000 & 4.330 & 1.048 & -1.778 & 2.296 \\ 
&  & (0.080) & (0.051) & (0.388) & (0.081) \\ 
0.60 & 0.000 & 4.412 & 1.045 & -1.767 & 2.296 \\ 
&  & (0.081) & (0.049) & (0.325) & (0.073) \\ 
0.70 & 0.001 & 4.506 & 1.038 & -1.756 & 2.296 \\ 
&  & (0.086) & (0.049) & (0.326) & (0.070) \\ 
0.80 & 0.001 & 4.633 & 1.020 & -1.723 & 2.296 \\ 
 &  & (0.111) & (0.060) & (0.304) & (0.071) \\ 
0.90 & 0.002 & 4.850 & 0.982 & -1.679 & 2.296 \\ 
&  & (0.158) & (0.086) & (0.325) & (0.101) \\ 
0.95 & 0.001 & 5.100 & 0.929 & -1.675 & 2.296 \\ 
&  & (0.269) & (0.156) & (0.358) & (0.124) \\ 
0.98 & 0.002 & 5.484 & 0.841 & -1.648 & 2.276 \\ 
&  & (0.303) & (0.196) & (0.264) & (0.136) \\ 
0.99 & 0.006 & 5.631 & 0.872 & -1.626 & 2.153 \\ 
 &  & (0.255) & (0.157) & (0.202) & (0.107) \\ 
   \hline
\end{tabular}
}
\end{table}

\section{Concluding Remarks} 
\label{s:conclude}

In this article, we have developed the continuous threshold expectile regression model. 
This model allows the expectiles of the response to be piecewise linear but still continuous in covariates. 
We developed a grid search method to estimate the unknown threshold and the regression coefficients.  
A weighted CUSUM type test statistics was proposed to test the structural change 
at  a given expectile.  
Our numerical studies showed that the proposed estimator has good finite sample performance. 

Our work may be extended in several ways. 
First, although generally there are fewer crossings in expectile regression than in quantile regression \citep{schnabel2009optimal},  the expectile crossings may happen.    
It will be worthwhile to extend our model to  non-crossing continuous threshold expectile estimation and to develop tests for structure change across expectiles.  
Another interesting extension is to consider more than one threshold for a covariate. 
In such a situation, the estimation and test of the thresholds would be more complicated,  
and further investigation is needed.  

\section*{Acknowledgements}
The authors thank  Dr. Andrew Wiesner for the interpretation of the baseball data. 
This research is partially supported by NIH R01GM109453. 
Zhang's work is partially supported by National Natural Science Foundation of China (NSFC) 
(11401194), and the Fundamental Research Funds for the Central Universities (531107050739).


\begin{appendices}
\section{}
\underline{Regularity Conditions}. 
\begin{description}
	\item[(A1)] 
	$t_0=\arg\min_{t\in \mathcal{T}} M_{\tau}\left(\widehat{\bxi}(t), t\right)$  is unique, 
	where $\mathcal{T}$ is a compact set in $\mathbb{R}^1$. 
	
	\item[(A2)] 
	$\btheta_\tau$ is in $\Theta$, and $\Theta$ is a compact subset of $\mathbb{R}^{p+4}$. 
	
	\item[(A3)] 
	 The scalar variable $X$ has an absolutely continuous distribution with density function $f_X$, 
	 which	is strictly positive, bounded and continuous for any $t$ in a neighborhood of $t_0$. 
	 
	 \item[(A4)] 
	 $\mbox{E}|Y|^2<\infty$, $\mbox{E}|X|^2<\infty$,  and $\mbox{E}|\bZ|^2<\infty$. 
	 
	 \item[(A5)]  
	 Given $\beta_2\neq 0$, the Hessian matrix  $H(\btheta_0)$ is nonsingular. 
\end{description}
Condition (A1) is the identifiability condition of the estimation.  
Conditions (A1)---(A3) are for the consistency of the estimates, and Conditions (A4)---(A5) are used for the asymptotic normality. 

We first provide the following  uniformly convergence results.  
\begin{lemma}\label{lmm1}
	Under the regular conditions,  as $n\rightarrow \infty$,  we have 
	$$
	\sup_{\btheta\in \Theta}\left|M_{n,\tau}(\btheta)-M_\tau(\btheta)\right|  \mathop{\longrightarrow}^{P}0.
	$$ 
\end{lemma}

\noindent
\textit{\underline{Proof of Lemma \ref{lmm1}}}.  
To show that the class of functions $\{m_{\btheta}: \btheta\in \Theta\}$ is Glivenko-Cantelli, 
it is sufficient to show $m_{\btheta}$ is Lipschitz continuous. 
Recalling that $\btheta=(\bxi^\top, t)^\top$, and the derivatives 
\begin{align*}
	\frac{\partial m_{\btheta}}{\partial\bxi}
	&=-2\omega_\tau \bV(t) \left[Y- \bxi^\top \bV(t)\right],\\
	\frac{\partial m_{\btheta}}{\partial t}
	& =2 \omega_\tau \beta_2 I(X>t)\left[Y- \bxi^\top \bV(t)\right].
\end{align*}
By the Condition (A2),  
both $|\max \bV(t)\left[Y-\bxi^\top \bV(t)\right]| $ and $\max|\beta_2 I(X>t)|$ are finite.  
Note that $w_\tau\leq \max(\tau, 1-\tau)<1$ for any $\tau\in (0,1)$.  
Hence, applying the mean-value theorem, 
$\left|m_{\btheta_1}(\mathcal{X}) - m_{\btheta_2}(\mathcal{X})\right|\leq m(\mathcal{X})\|\btheta_1-\btheta_2\|$ for every $\mathcal{X}$,  
where 
\[
m(\mathcal{X})=
  \begin{bmatrix}
   \max |\bV(t)\left[Y-\bxi^\top \bV(t)\right]|  & \\
   \max|\beta_2 I(X>t) \left[Y- \bxi^\top \bV(t)\right] |& 
  \end{bmatrix}
  <   \infty.
\]
Therefore, $m_{\btheta}$ is Lipschitz continuous,  and applying the Glivenko-Cantelli theorem and Example 19.8 in \citet{van2000asymptotic},  
we can establish that $\{m_{\btheta}: \btheta\in \Theta\}$ is Glivenko-Cantelli.  \qed

\medskip
\noindent
\textit{\underline {Proof of Theorem \ref{thm1}}}.  
By Lemma \ref{lmm1},  $\sup_{\btheta\in \Theta}\left|M_{n,\tau}(\btheta)-M_\tau(\btheta)\right|  \mathop{\longrightarrow}\limits^{P}0$ as $n$ goes to infinity. 
Since $\Theta$ is compact and the uniqueness of the minimum $\btheta_0$ (by Conditions A1 and A2), along with that $M_{n,\tau}(\btheta)$ is continuous with respective to $\btheta$, then we can establish that  $\widehat{\btheta}\mathop{\longrightarrow}\limits^P \btheta$,  
by Theorem 2.1 of \citet{newey1994large}.  \qed

\medskip
\noindent
\textit{\underline {Proof of Theorem \ref{thm2}}}.  
Firstly,  
by Condition (A3),  
the function $\mathcal{X}\longmapsto m_{\btheta}(\mathcal{X})$ is measurable, 
and the function $\btheta \longmapsto m_{\btheta}(\mathcal{X})$ is differentable at $\btheta_0$ for P-almost every $\mathcal{X}$. 
Recall that $m_{\btheta}$ is Lipschitz continuous with respect to $\btheta$, as proved in Lemma \ref{lmm1}. 

Secondly, the map $\btheta\longmapsto M_\tau(\btheta)= \mbox{E}m_{\btheta}$ admits a second order Taylor expansion at $\btheta_0$, with a nonsingular symmetric Hessian matrix 
$H(\btheta_0)$. 
We can verify that $H(\btheta)$ is continuous in $\btheta$.  Indeed, the elements of $H(\btheta)$ are quadratic functions of $\bxi$, and hence $H(\btheta)$ is continuous in $\bxi$.  
It is sufficient to show that $H(\btheta)$ is continuous in $t$.  Note that the first term of $H(\btheta)$ is a function of $t$ through moments of the form  
$\mbox{E}\left[\bV(t)I(X>t)\right]$. 
By Condition (A4),  $(\mbox{E}(\|\bV(t))\|^2)^{1/2}\leq C_1$ for some constant $C_1<\infty$.  
By Condition (A3),  $|F_X(t_1)-F_X(t_2)|\leq \max_x f_X(x) |t_2-t_1|\leq C_2 |t_2-t_1|$ for some constant $C_2<\infty$, $t_1 < t_2$. Then, by Cauchy-Schwartz inequality, 
\begin{align*}
	\mbox{E}\| \bV(t)I(t_1\leq X \leq t_2)\|^2
	&\leq \left(\mbox{E}\| \bV(t)\|^2\right)^{1/2} 
	\left(\mbox{E}|t_1\leq X \leq t_2|^2\right)^{1/2} \\
	& \leq C_1C_2|t_1-t_2|^{1/2}, 
\end{align*}
is uniformly continuous in $t$.  
Hence, the first term of $H(\btheta)$ is continuous in $t$. 
On the other hand, since $\mbox{E}\omega_\tau=\tau\left(1-F_Y(\bxi^\top \bV(t))\right)+(1-\tau)F_Y(\bxi^\top \bV(t))$ is continuous in $t$,  then the second term of $H(\btheta)$ is continuous in $t$.  Thus,  $H(\btheta)$ is continuous in $t$. 

Finally, by Theorem \ref{thm1}, $\widehat{\btheta}$ is consistent for $\btheta_0$ in a neighborhood of $\btheta_0$,  
it follows that $\sqrt{n}(\widehat{\btheta}-\btheta_0)$ is asymptotically normal with mean zero and covariance matrix $H(\btheta_0)^{-1}\Sigma(\btheta_0) H(\btheta_0)^{-1}$,  
by Theorem 5.23 in  \citet{van2000asymptotic}.    
\qed

\begin{lemma}\label{lmm2}
	Under the regularity conditions,  as $n\rightarrow \infty$,  we have 
	\begin{description}
	\item[(i)] $\widehat{S}_{wn}(\widehat{\balpha})\mathop{\longrightarrow}\limits^P
	S_w(\balpha)$.

	\item[(ii)] 
	$\sup_t \left|\widehat{S}_{1n}(\widehat{\balpha}, t)- S_1(\balpha, t)\right|
	\mathop{\longrightarrow}\limits^P 0$.

	\item[(iii)] 
	$\sup_t \left|\widehat{S}_{2n}(\widehat{\balpha}, t)- S_2(\balpha, t)\right|
	\mathop{\longrightarrow}\limits^P 0$.	
\end{description}
\end{lemma}
\medskip
\noindent
\textit{\underline {Proof of Lemma \ref{lmm2}}}.  
It is easily obtained by using the law of large number for (i).  
To establish (ii) and (iii), note that $\widehat{S}_{1n}(\widehat{\balpha}, t)$ and 
$\widehat{S}_{1n}(\widehat{\balpha}, t)$ are sums of indicator functions and Lipschitz functions, then they are Glivenko-Cantelli class, which implies that both (ii) and (iii) holds.

\medskip
\noindent
\textit{\underline {Proof of Theorem \ref{thm3}}}.  
Note that 
$\widehat{\balpha}=\arg\min_{\balpha}n^{-1/2}\sum_{i=1}^n 
\left|\tau-I(Y_i\leq \balpha^\top \bW_i)\right|(Y_i- \balpha^\top \bW_i)^2$,  
which is equivalent to the solution of the estimating equation 
\[
	U_n(\balpha)=\frac{1}{\sqrt{n}}\sum_{i=1}^n \left|\tau-I(Y_i\leq \balpha^\top \bW_i)\right|
	\bW_i(Y_i-\balpha^\top \bW_i).
\]
Recall that  the local alternative model \eqref{mod2} is 
\[
	Y_i = \beta_0 + \beta_1 X_i + n^{-1/2}\beta_2 (X-t)_+ + \bgamma^\top \bZ + e_i. 
\]
Then, under model  \eqref{mod2},  the estimating equation can be written as 
\begin{align*}
	U_n(\balpha)=\frac{1}{\sqrt{n}}\sum_{i=1}^n \left|\tau-I(Y_i\leq \balpha^\top \bW_i)\right|
	\bW_ie_i + \frac{1}{n}\sum_{i=1}^n \left|\tau-I(Y_i\leq \balpha^\top \bW_i)\right|
	\bW_i \beta_2(X_i-t) I(X_i>t) + o_P(1).
\end{align*}
By the mean-value theorem,  we have 
\begin{align*}
	-U_n(\balpha) 
	&= U_n(\widehat{\balpha})-U_n(\balpha) \\
	&= -\frac{1}{\sqrt{n}}\sum_{i=1}^n 
	\left|\tau-I(Y_i\leq \widehat{\balpha^*}^\top \bW_i)\right|
	\bW_i \bW_i^\top (\widehat{\balpha}-\balpha) +o_p(1)\\
	&= - \widehat{S}_{wn}(\widehat{\balpha^*}) \sqrt{n} (\widehat{\balpha}-\balpha)+o_p(1). 
\end{align*}
where $\widehat{\balpha^*}$ lies in the line between $\widehat{\balpha}$ and $\balpha$.  
By Lemma \ref{lmm2},  
$\widehat{S}_{wn}(\widehat{\balpha})\mathop{\longrightarrow}\limits^PS_w(\balpha)$,   
and under the local alternative model \ref{mod2},  
it yields that  
\begin{align*}
	\sqrt{n}(\widehat{\balpha}-\balpha)
	&= \frac{1}{\sqrt{n}}S_w(\balpha)^{-1}\sum_{i=1}^n 
	\left|\tau-I(Y_i\leq \balpha^\top \bW_i)\right|
	\bW_i (Y_i-\balpha^\top \bW_i)+o_P(1)\\
	&=\frac{1}{\sqrt{n}}S_w(\balpha)^{-1}\sum_{i=1}^n 
	\left|\tau-I(Y_i\leq \balpha^\top \bW_i)\right|
	\bW_ie_i  \\
	&+ \frac{1}{\sqrt{n}}S_w(\balpha)^{-1}\sum_{i=1}^n 
	\left|\tau-I(Y_i\leq \balpha^\top \bW_i)\right| \bW_i \beta_2 (X_i-t)I(X_i>t)+o_P(1).
\end{align*}
Thus,  under the local alternative model \ref{mod2}, by plugging in the representation for
$\sqrt{n}(\widehat{\balpha}-\balpha)$ and some algebraic manipulation, we have
\begin{align*}
	R_n(t)
	&= \frac{1}{\sqrt{n}}\sum_{i=1}^n 
	\left|\tau-I(Y_i\leq \balpha^\top \bW_i)\right|
	\bigg[Y_i-\balpha^\top \bW_i-n^{-1/2}\beta_2 (X_i-t)I(X_i>t)\\
	&-(\widehat{\balpha}-\balpha)^\top\bW_i +n^{-1/2}\beta_2 (X_i-t)I(X_i>t) 
	\bigg](X_i-t)I(X_i\leq t) + o_P(1)\\
	&=  \frac{1}{\sqrt{n}}\sum_{i=1}^n 
	\left|\tau-I(Y_i\leq \balpha^\top \bW_i)\right|e_i \left[(X_i-t)I(X_i\leq t)
	-\widehat{S}_{1n}(\balpha, t)^\top \widehat{S}_{wn}(\balpha)^{-1}\bW_i\right]\\
	&- \widehat{S}_{1n}(\balpha, t)^\top \widehat{S}_{wn}(\balpha)^{-1} 
	\widehat{S}_{2n}(t, \balpha) + o_P(1)\\
	&=\frac{1}{\sqrt{n}}\sum_{i=1}^n 
	\left|\tau-I(Y_i\leq \balpha^\top \bW_i)\right|e_i \left[(X_i-t)I(X_i\leq t)
	-\widehat{S}_{1n}(\balpha, t)^\top \widehat{S}_{wn}(\balpha)^{-1}\bW_i\right]
	-q(t) + o_P(1).
\end{align*}
It is easy to derive the remainder conclusion for weak convergence of  $R_n(\widehat{\balpha}, t)$
by following the proofs in \citet{stute1997nonparametric}. \qed

\medskip
\noindent 
\textbf{Proof of Theorem~\ref{thm3}}

We divide the proof into three steps. 
Firstly, we show that the covariance function of $R_n^*$ converges to that of $R$. Define 
\begin{align*}
	R_n^*( t)
	&= \frac{1}{\sqrt{n}}\sum_{i=1}^{n}  v_i
				(Y_i-\balpha^\top \bW_i)
				\left|\tau-I(Y_i\leq\balpha^\top \bW_i)\right|\\
				& \times
	\left[
	(X_i-t)I(X_i\leq t)- 
	S_{1}(\balpha, t)^\top S_{w}(\balpha)^{-1}W_i
	\right]. 
\end{align*}
By the fact that the consistency of $\widehat{\balpha}-\balpha$,  
along with the uniform convergence of $\widehat{S}_{1n}(\widehat{\balpha}, t)- S_1(\balpha, t)$ and $\widehat{S}_{wn}(\widehat{\balpha})-S_w(\balpha)$, 
one can easily show $R_n^*(t)$ and $R_n^{**}(t)$ are asymptotically equivalent in the sense that 
\[
	\sup_t \|R_n^*(t)-R_n^{**}(t)\|=o_P(1).
\]
Note that $ v_i$'s are independent of $(Y_i,\bX_i, Z_i)$,  
and $\mbox{E}v_i=0$, $\mbox{Var}(v_i)=1$.   
Then, for any $t_1, t_2$, 
the covariance function of $R_n^{**}$ is 
\begin{eqnarray*}
  &&	
  Cov\left(R_n^{**}(t_1), R_n^{**}(t_2)\right)\\
  &=&
  \frac{1}{n}\sum_{i=1}^{n}
 \mbox{E}\bigg(v_i^2
 e_i^2
  \left|\tau-I(e_i\leq 0)\right|^2
  \left\{
  (X_i-t_1)I(X_i\leq t_1)  - S_1(\balpha, t_1)^TS_w(\balpha)^{-1}\bW_i
  \right\}\\
  && \times
  \left\{
 (X_i-t_2)I(X_i\leq t_2) - S_1(\balpha, t_2)^TS_w(\balpha)^{-1}\bW_i
  \right\}\bigg)\\
  &=&  
  \mbox{E}
  	\bigg[
  	e^2
  \left|\tau-I(e\leq 0)\right|^2
  \left\{
  (X-t_1)I(X\leq t_1)  - S_1(\balpha, t_1)^TS_w(\balpha)^{-1}\bW
  \right\}\\
  &&\times 
  \left\{
  (X-t_2)I(X\leq t_2)  - S_1(\balpha, t_2)^TS_w(\balpha)^{-1}\bW
  \right\}
  	\bigg]. 
\end{eqnarray*}
which  is the same as the covariance of $R(t)$. 

Secondly, it is easily to show that any finite-dimensional projection of $R_n^{**}(t)$ converges to that of $R(t)$,  by the central limit theorem.

Thirdly,  $R_n^{**}(t)$ is uniformly tight.  
Note that the class of all indicator functions $I(X\leq t)$ is a  Vapnik-Chervonenskis (VC) class of functions. 
Then, the class of functions 
$$
\mathcal{F}_n=\left\{
(X_i-t)I(X_i\leq t)- 
S_{1n}(t)S_w^{-1}\bW_i: 
t \in \mathbb{R}^1 
\right\}
$$ 
is a VC class of functions. Thus, by the equicontinuity lemma~15 of \citep{pollard1984convergence}, 
one can show that $R_n^*(\tau)$ is uniformly tight. 
Then, by the Cramer-Wold device, the proof of Theorem~2.3 
is completed. 
\qed

\end{appendices}

\section*{References}
\bibliographystyle{elsarticle-harv} 
\bibliography{myreference}





\end{document}